\documentclass[prb,preprint]{revtex4-1}

\usepackage{amsmath}  
\usepackage{amsfonts} 
\usepackage{graphicx} 
\usepackage{tensor}
\usepackage{cancel}
\usepackage{units}
\usepackage{caption2}
\usepackage{txfonts}
\usepackage{bbm}
\usepackage{cancel}
\usepackage{listings}
\usepackage{comment}
\begin{document}
\title{ A method to calculate inductance in systems of parallel wires}

\author{Eric Deyo}
\email{ecdeyo@fhsu.edu}
\affiliation{Department of Physics, Fort Hays State University, Hays, KS 67601}

\date{\today}
\begin{abstract}
This paper gives a method that maps the static magnetic field due to a system of parallel current-carrying wires to a complex function.  Using this function simplifies the calculation of the magnetic field energy density and inductance per length in the wires, and we reproduce well-known results for this case.
\end{abstract}
\maketitle
This paper points out a method based on complex analysis \cite{Whittaker} that may yield a simplification in the calculation of the inductance of a system consisting of parallel wires.  The method is an interesting example of applying complex analysis in its own right, and reproduces well-known results \cite{Landau, Jackson}.  There are many papers using similar techniques using stream functions in fluids to understand vortices \cite{Lamb}, and also some magnetic field problems, mostly in the absence of currents \cite{Stratton}.

Ampere's law in SI units in integral form is
\begin{equation}
\oint_c{\bf B}\cdot d{\bf r} = \mu_0 I_{enc},
\end{equation}
where $I_{enc}$ is the current puncturing the area enclosed by path $c$, with the sign of the current given by the right hand rule.

If the path $c$ is in the $x$-$y$ plane, Ampere's law formally looks like
\begin{equation}
\label{Ampere2}
\oint_c \left(B_x dx +B_y dy\right) = \mu_0 I,
\end{equation}
with positive currents $I$ moving in the positive $z$-direction.

Consider a complex function $f(z)$ of $z=x+iy$, such that $f(z) = B_x -iB_y$.  Then substituting $f$ into the contour integral $\int_c f(z) dz$, where the contour $c$ in the complex plane is identical to the path of the line integral in \eqref{Ampere2} (meaning the same x- and y-values are traversed in the line integral as in the complex contour integral), yields
\begin{equation}
\label{complexintegral}
\int_c f(z)dz = \oint_c\left( B_x dx +B_y dy\right) + i \oint_c\left( B_x dy -B_y dx\right).
\end{equation}
Therefore,
\begin{displaymath}
\oint_c{\bf B}\cdot d{\bf r} = Re\Bigg\{\int_c f(z)dz\Bigg\} = \mu_0 I_{enc}.
\end{displaymath}
Now let's take a look at the imaginary part of the complex integral \eqref{complexintegral}.  This is
\begin{displaymath}
\oint_c\left( B_x dy -B_y dx\right).
\end{displaymath}
This can be written in terms of a parameter $t$ that parameterizes the contour as
\begin{displaymath}
\int_{t_1}^{t_2} \left(B_x \frac{dy}{dt} -B_y \frac{dx}{dt}\right) dt.
\end{displaymath}
The integrand is formally $B_xv_y-B_yv_x = -\left({\bf v}\times {\bf B}\right)_z$.  This is proportional to the z-component of the Lorentz force on a charge.  Now, let's suppose a test charge moves in a small circle of radius $r$, with a constant speed $v$ in the x-y plane around a wire carrying constant current $I$ in the z-direction   The magnetic field from the wire points parallel or antiparallel to the motion of the particle, and therefore the force in the z-direction is zero.  The imaginary part of the complex integral \eqref{complexintegral} around the contour corresponding to this small circle vanishes.  Since the complex integral can be written as the sum of integrals around small contours around its poles, this means that the imaginary part of the complex integral should vanish.

We are left with
\begin{equation}
\label{intf}
\int_c f(z)dz = \oint_c {\bf B}\cdot d{\bf r} = \mu_0 I
\end{equation}
as our condition on $f$ to be $B_x-iB_y$.
Hence, since by the residue theorem,
\begin{displaymath}
\int_c f(z)dz = 2\pi i R,
\end{displaymath}
where $R$ is the sum of the residues of $f$ enclosed in $c$, we have that
\begin{equation}
\label{R1}
2 \pi i R = \mu_0 I.
\end{equation}

A simple form of $f$ which satisfies \eqref{intf} for a wire carrying current I in the z-direction at $(x_0, y_0)$ is
\begin{equation}
\label{f1}
f\left(z\right) = \frac{\mu_0 I}{2\pi i \left(z - z_0\right)},
\end{equation}
where $z_0 = x_0+iy_0$.
Taking the real and imaginary parts of $f(z)$, and comparing with $f = B_x-iB_y$, we find
\begin{eqnarray}
B_x = -\frac{\mu_0 I}{2\pi r} \frac{y-y_0}{r}, \nonumber\\
B_y =   \frac{\mu_0 I}{2\pi r}\frac{x-x_0}{r},
\end{eqnarray}
where $r^2 = (x-x_0)^2 + (y-y_0)^2$.
This is precisely the field ${\bf B} = \frac{\mu_0 I}{2\pi r} \hat{\bf \phi}$ that one expects according to Ampere's law.

The reader can easily see that \eqref{f1} is not unique in yielding an integral whose residue obeys \eqref{R1}.  In fact adding any analytic function to \eqref{f1} will give an identical result and the condition $\nabla\cdot {\bf B}$ will hold as well. For the purpose of this paper, we will choose \eqref{f1} because it yields the expected field for a single wire.
Now, it is straightforward to generalize \eqref{f1} to any number of wires carrying current in the z-direction, at the positions $(x_j,y_j)$.  We simply write
\begin{equation}
\label{f}
f(z)  = \frac{\mu_0}{2\pi i} \sum_j \frac{I_j}{z-z_j},
\end{equation}
and then take the real and imaginary parts of this to find the $x$ and $y$ components of the magnetic field.

Now, the energy density in a magnetic field is $u_B = \frac{1}{2\mu_0} B^2$.  So for the magnetic field in the x- and y-directions, we can write
$u_B = \frac{1}{2\mu_0}f^*f$.  The total energy stored in the magnetic field is then
\begin{equation}
\label{energy}
U = \frac{1}{2\mu_0} \int f^*f d^3{\bf r}.
\end{equation}
The total energy stored in a magnetic field that is created by a system of currents $I_j$ is related to the inductances according to
\begin{equation}
\label{inductancedef}
U = \frac{1}{2} \sum_{j,k}L_{jk}I_jI_k
\end{equation}
where $I_j$ is the current in wire $j$, and $L_{jk}$ is symmetric in its indices, and is the mutual inductance of wires $j$ and $k$.  $L_{kk}$ is the self inductance of wire $k$.

Suppose we add a current $I_n$ to the system of currents in the z-direction.
Let $U_0$ be the energy stored in the magnetic field in the absence of current $I_n$ and let $U$ be the energy stored in the magnetic field in the presence of $I_n$.  We can then write
\begin{equation}
\label{deltaUoverI}
\frac{U-U_0}{I_n} = \frac{1}{2}\sum_{j\neq n}\left[L_{nj}I_j +L_{jn}I_j\right]+\frac{1}{2}L_{nn} I_n = \sum_{j\neq n} L_{nj}I_j + \frac{1}{2} L_{nn} I_n
\end{equation}
Let $f_0$ be the function $f(z)$ in the absence of $I_n$ and let $f$ be the function $f(z)$ in the presence of $I_n$.  Then
\begin{equation}
\label{energychange}
U-U_0 = \frac{1}{2\mu_0}\int \left[|f|^2-|f_0|^2\right] d^3{\bf r}
\end{equation}
and
\begin{equation}
\label{fchange}
f = f_0 + \frac{\mu_0 I_n}{2 \pi i \left(z-z_n\right)}.
\end{equation}
Plugging \eqref{fchange} into \eqref{energychange}, then dividing by $I_n$, we find that
\begin{equation}
\label{deltaUoverI2}
\frac{U-U_0}{I_n} = \frac{1}{4\pi i} \int \left[ \frac{f_0^*}{z-z_n} -\frac{f_0}{z^*-z_n^*} +\frac{iI_n}{2\pi |z-z_n|^2}\right]d^3{\bf r}
\end{equation}
Plugging in the definition of $f$ from \eqref{f}, we find that
\begin{equation}
\label{deltaUoverI3}
\frac{U-U_0}{I_n}= \sum_{j \neq n} I_j \frac{\mu_0}{8 \pi^2} \int\left[\frac{1}{(z-z_j)(z^*-z_n^*)} + c.c.\right] d^3{\bf r} + I_n \frac{\mu_0}{8 \pi^2}\int \frac{1}{(z-z_n)(z^*-z_n^*)} d^3{\bf r}
\end{equation}
Comparing \eqref{deltaUoverI3} with \eqref{deltaUoverI}, we arrive at a formulae for both the self inductance and the mutual inductance.  The self inductance is
\begin{equation}
\label{selfinductance}
L_{nn} = \frac{\mu_0}{4\pi^2}\int\frac{1}{(z-z_n)(z^*-z_n^*)}d^3{\bf r}= \frac{\mu_0l}{4\pi^2}\int\frac{1}{(z-z_n)(z^*-z_n^*)}dxdy,
\end{equation}
and the mutual inductance is
\begin{eqnarray}
\label{mutualinductance}
L_{nj} &=& \frac{\mu_0}{8\pi^2}\int\left[\frac{1}{(z-z_j)(z^*-z_n^*)}+\frac{1}{(z^*-z_j^*)(z-z_n)}\right]d^3{\bf r} \nonumber\\
&=&\frac{l\mu_0}{8\pi^2}\int\left[\frac{1}{(z-z_j)(z^*-z_n^*)}+\frac{1}{(z^*-z_j^*)(z-z_n)}\right]dxdy.
\end{eqnarray}

The self-inductance, equation \eqref{selfinductance}, can be directly integrated.  We note that $(z-z_n)(z^*-z_n^*) = (x-x_n)^2+(y-y_n)^2 = r^2$.  Where we set the origin of the coordinate system to the position of wire $n$ and then converted to polar coordinates.
Performing the integration in polar coordinates, we arrive at the inductance per length of the wire being
\begin{equation}
\label{selfinductance2}
\frac{L_{nn}}{l} = \frac{\mu_0}{4\pi^2}2\pi\int_a^{\Lambda} \frac{1}{r^2}r dr = \frac{\mu_0}{2\pi} \ln\left(\frac{\Lambda}{a}\right).
\end{equation}
Here we introduced long range and short range cutoffs for the integration, $\Lambda$ and $a$ respectively, and the self-inductance is only written to logarithmic accuracy, as per usual.

The integral for the mutual inductance can also be done, but is a little more involved.  Here again, we find it helpful to set the origin of the $x-y$ coordinate system to the position of wire $n$, and then convert to polar coordinates. Let $z = re^{i\theta}$ and $z_j = r_j e^{i\phi}$.  Then the integral \eqref{mutualinductance} becomes
\begin{equation}
\label{mutualinductance2}
\frac{L_{nj}}{l} = \frac{\mu_0}{8\pi^2}\int_0^\infty dr\int_0^{2\pi} d\theta \left[\frac{1}{r-r_je^{-i(\phi-\theta)}}+
\frac{1}{r-r_je^{i(\phi-\theta)}}\right].
\end{equation}
We first perform the integral over $\theta$.  We do this via residues. In the first integral on the right hand side of \eqref{mutualinductance2}, write $u = e^{-i\theta}$, and then $d\theta = \frac{du}{-iu}$, and the integral becomes a contour integral over a unit circle in the complex-$u$ plane, traversed in the clockwise direction, call this contour $-c$.  In the second integral on the right hand side, we write $u = e^{i\theta}$, so $d\theta =\frac{du}{iu}$, and the integral becomes the integral over  a unit circle in the complex-u plane, traversed in the counterclockwise direction.  We'll call this contour $c$.  Performing these contour integrals, we find that the integral over $\theta$ in \eqref{mutualinductance2} is
\begin{equation}
\label{angleintegral}
\int_0^{2\pi} d\theta \left[\frac{1}{r-r_je^{-i(\phi-\theta)}}+
\frac{1}{r-r_je^{i(\phi-\theta)}}\right] =
\Bigg\{
\begin{array}{ccc}
&\frac{4\pi}{r} \qquad r> r_j \\
&0 \qquad r<r_j
\end{array}
\end{equation}
 In place of $r_j$, for generalization purpose, we introduce the distance $r_{nj}$ which is the distance between wires $n$ and $j$.  The mutual inductance per length is given by
\begin{equation}
\label{mutualinductance3}
\frac{L_{nj}}{l} = \frac{\mu_0}{2\pi} \int_{r_{nj}}^\Lambda \frac{dr}{r} = \frac{\mu_0}{2\pi}\ln\left(\frac{\Lambda}{r_{nj}}\right),
\end{equation}
where we introduced a long range cutoff $\Lambda$ for the integral over $r$.  Again this is written only to logarithmic accuracy.

We further note that these results are well known, though our method is different.  In a future paper, we hope to apply this formalism to the calculation of inductance in different systems.  We would also like to mention that the similarity between our formalism and the velocity stream function in fluid flow, with currents being replaced by vorticity \cite{Lamb} lead naturally to a nice qualitative picture of the magnetic field around arrays of wires, or around arrays of currents.   In the future, we would to examine the interplay between the energy density of currents in a solid and the attraction between parallel current carrying wires.  It is our belief that the magnetic field may break into an array vortices similar in structure to the currents around magnetic flux lines in type II superconductors, depending on the solid \cite{Abrikosov, Feynman}.

\begin{acknowledgments}
The author gratefully acknowledges Luis Pauyac and Emma Diextre for helpful conversations about this subject.
\end{acknowledgments}

\end{document}